\documentclass[doublecol,linenumbers]{epl2} 

\usepackage{color}
\newcommand{\michal}[1]{\textcolor{black}{#1}}
\newcommand{\michall}[1]{\textcolor{black}{#1}}

\begin{document}

\title{Effect of boundary conditions in turbulent thermal convection}
\shorttitle{Effect of boundary conditions in turbulent thermal convection} 

\author{P. Urban\inst{1} \and T. Kr\'{a}l\'{i}k\inst{1} \and M. Macek\inst{1} \and  P. Hanzelka\inst{1}  \and T. V\v{e}\v{z}n\'{i}k\inst{1} \and L. Skrbek\inst{2}}
\shortauthor{P. Urban \etal}

\institute{                    
  \inst{1} The Czech Academy of Sciences, Institute of Scientific Instruments, Kr\'{a}lovopolsk\'{a} 147, Brno, Czech Republic\\
  \inst{2} Faculty of Mathematics and Physics, Charles University, Ke Karlovu 3, Prague, Czech Republic
}
\pacs{47.27.-i}{Turbulent flows}
\pacs{47.55.pb}{Thermal convection}
\pacs{47.27.te}{Turbulent convective heat transfer}

\abstract{We report an experimental study aiming to clarify the role of boundary conditions (BC) in high Rayleigh number $10^8 < {\rm{Ra}} < 3 \times 10^{12}$ turbulent thermal convection of cryogenic helium gas. We switch between BC closer to constant heat flux (CF) and constant temperature (CT) applied to the highly conducting bottom plate of the aspect ratio one cylindrical cell 30~cm in size, leading to dramatic changes in the temperature probability density function and in power spectral density of the temperature fluctuations measured at the bottom plate, while the dynamic thermal behaviour of the top plate and bulk convective flow remain unaffected. Within our experimental accuracy, \michall{ we find no appreciable changes in Reynolds number Re(Ra) scaling, in the dimensionless heat transfer efficiency expressed via Nusselt number Nu(Ra) scaling, nor in the rate of direction reversals of large scale circulation}. }

\maketitle

\section{Introduction}
Confined 3D turbulent Rayleigh-B\'{e}nard convection (RBC)~\cite{AhlersRevModPhys,ChillaSchumacher} serves as a 
model system on our way to understand many natural phenomena and represents the most frequently studied thermally driven turbulent flow.
The ideal laterally infinite RBC occurs in a fluid layer confined between two horizontal, perfectly conducting plates heated from below in a gravitational field, and for an Oberbeck-Boussinesq (OB) fluid it is fully characterized by the Rayleigh (Ra) and the Prandtl (Pr) numbers. The convective heat transfer efficiency is described by the Nusselt number, via the Nu = Nu(Ra; Pr) dependence. Large scale circulation (LSC), also known as ‘wind’ \cite{WindJFM} of mean velocity $U$ and dimension of the size of the convective layer, $L$ (or the size $D$ of the RBC cell in the case of laterally confined RBC) is known to exist in RBC and can be characterized by the Reynolds number (Re). The dimensionless numbers describing confined RBC are defined as:
\begin{equation}
\label{numbersDef}
 {\rm{Nu}} = \frac{Lq}{\lambda_f \Delta T};\    {\rm{Ra}}=\frac{g\alpha_f}{\nu_f\kappa_f}\Delta T L^3; \  {\rm{Pr}}=\frac{\nu_f}{\kappa_f}; \   {\rm{Re}} = \frac{UL}{\nu_f}.
\end{equation}
Here $q$ is the total convective heat flux density, $g$ stands for the acceleration due to gravity, and $\Delta T$ is the temperature difference between the parallel top and bottom plates separated by the vertical distance $L$. The properties of the working fluid are characterized by the thermal conductivity, $\lambda_f$, and by the combination $\alpha_f/(\nu_f \kappa_f)$, where $\alpha_f$  is the isobaric thermal expansion, $\nu_f$  is the kinematic viscosity, and $\kappa_f$ denotes the thermal diffusivity; $\kappa_f=\lambda_f/(\rho_f c_{pf})$, where $\rho_f$ is the density and $c_{pf}$ specific heat of the working fluid at constant pressure. In the laboratory the RBC experiments often take place in cylindrical cells of diameter $D$ and height $L$; the relevant additional parameter is the aspect ratio defined as $\Gamma = D/L$, which might be understood as a first approximation in taking into account the shape of the RBC cell.

\begin{table*} [t]
\caption{Experimental quantities relevant for generation of cryogenic turbulent RBC flows in the Brno experimental cell \cite{BrnoCellRSI} in comparison with hypothetical RBC flows of the same Ra and Nu (assuming the same Nu=Nu(Ra) scaling) that would be requested at ambient temperatures using $\rm{H_2O}$ and $\rm{SF}_6$ as working fluids~\cite{McCartyArp,H2O,SF6}. The $\Gamma=1$ Brno cell has $e=28$~mm thick top and bottom plates 30 cm in diameter made of annealed copper of thermal conductivity $\lambda_p$=2210 and 400  W~m$^{-1}$K$^{-1}$ and thermal capacity $c_p=0.144$ and 386 J~kg$^{-1}$K$^{-1}$ at $T_m=(T_t+T_b)/2 \approx 5$~K and 300~K, respectively~\cite{CryoComp}. The influence of the vertical wall, made of nominally 0.5 mm thick stainless steel, is neglected. For definition of the displayed physical quantities $\alpha_f \Delta T, Q_b, \frac{\lambda_p}{\rm{Nu}\lambda_f}, \tau_p^h, \tau_f^h, \ell_p, K, \tau_{\rm{plm}}, \tau_{\rm{plt}}$, see the text.}
\label{tab.1}
\begin{center}
\begin{tabular}{lc|lll|lll|ll}
    &  unit & He & $\rm{H_2O}$ &  ${\rm{SF}_6}$ & He & $\rm{H_2O}$ & ${\rm{SF}_6}$ & He & ${\rm{SF}_6}$ \\
\hline 
\vspace{-0.3cm}
 & & & & & & & & & \\
Ra & 1 &$2.3{\rm x}10^{8}$&$2.3{\rm x}10^{8}$&$2.3{\rm x}10^{8}$&$1.3{\rm x}10^{10}$&$1.3{\rm x}10^{10}$&$1.3{\rm x}10^{10}$&$2.2{\rm x}10^{12}$&$2.2{\rm x}10^{12}$\\
Nu & 1 &41.7&41.7&41.7&138.3&138.3&138.3&717.8&717.8  \\
Pr & 1 &0.68&5.85 &0.79&0.71& 5.85 &0.79 & 1.00 &0.86 \\
\hline 
\vspace{-0.3cm}
 & & & & & & & & & \\
$T_m$  & K & 5.009 & 300.0 & 300.0 & 5.016 & 300.0 & 300.0 & 5.009 & 300.0\\
$\Delta T$  & K & 1.027 & 0.402 & 8.936 & 0.183 & 23.08 & 12.32 & 0.196 & 16.19\\
$P$ & Pa & $1.03{\rm x}10^{3}$ & $1.00{\rm x}10^{5}$ & $5.00{\rm x}10^5$ & $1.74{\rm x}10^{4}$ & $1.00{\rm x}10^5$ & $3.00{\rm x}10^5$ & $1.21{\rm x}10^5$ &  $1.8{\rm x}10^6$\\
$\nu_f$ &$\rm{\frac{m^2}{s}}$& $1.26{\rm x}10^{-5}$ & $8.57{\rm x}10^{-7}$ & $5.21{\rm x}10^{-6}$ & $7.4{\rm x}10^{-7}$ & $8.57{\rm x}10^{-7}$ &$8.46{\rm x}10^{-7}$ & $9.50{\rm x}10^{-8}$ &$1.2{\rm x}10^{-7}$ \\
$\kappa_f$ &$\rm{\frac{m^2}{s}}$& $1.85{\rm x}10^{-5}$ &  $1.46{\rm x}10^{-7}$ & $6.62{\rm x}10^{-6}$ & $1.04{\rm x}10^{-6}$ & $1.46{\rm x}10^{-7}$ & $1.08{\rm x}10^{-6}$ & $9.47{\rm x}10^{-8}$ &$1.4{\rm x}10^{-7}$ \\
$\alpha_f$ & 1/K & $2.00{\rm x}10^{-1}$ &$ 2.75{\rm x}10^{-4}$ & $3.40{\rm x}10^{-3}$ & $2.13{\rm x}10^{-1}$ &$ 2.75{\rm x}10^{-4}$ &  $3.74{\rm x}10^{-3}$ & $3.79{\rm x}10^{-1}$& $8.4{\rm x}10^{-3}$ \\
$\lambda_f$ &$\rm{\frac{W}{m~K}}$ & $9.57{\rm x}10^{-3}$ & $6.10{\rm x}10^{-1}$ & $1.30{\rm x}10^{-2}$ & $9.65{\rm x}10^{-3}$  & $6.10{\rm x}10^{-1}$ & $1.32{\rm x}10^{-2}$ & $1.05{\rm x}10^{-2}$ & $1.5{\rm x}10^{-2}$\\
\vspace{-0.34cm}
 & & & & & & & & & \\
$\rho_{f}$ & ${\rm{\frac{kg}{m^3}}}$ &0.10 & 996.6 & 2.94 & 1.72 & 996.6 & 18.19 & 15.03 & 139.2\\
\vspace{-0.34cm}
 & & & & & & & & & \\
$c_{pf}$ & $\rm{\frac{J}{kg~K}}$ & 5203 & 4181 & 667.9 & 5372 & 4181 & 673.5 & 7367 & 784.7\\
\hline
\vspace{-0.3cm}
 & & & & & & & & & \\
{\small$\alpha_f\Delta T$}& 1 &$2.10{\rm x}10^{-1}$ & $1.10{\rm x}10^{-4}$ & $3.00{\rm x}10^{-2}$ & $3.90{\rm x}10^{-2}$ & $6.30{\rm x}10^{-3}$ & $4.6{\rm x}10^{-2}$ & $7.4{\rm x}10^{-2}$ & $1.4{\rm x}10^{-1}$\\
$Q_b$ & W & 0.095 & 2.410 & 1.142 & 0.056 & 458.8 & 5.291 & 0.344 & 41.46\\
$\frac{\lambda_p}{{\rm{Nu}}\lambda_f}$& 1 & 5539 & 15.7 & 737 & 1656 & 4.7 & 219.3 & 293.8 & 36.8  \\
$\tau_p^h$ & s & 27.6 & 1140 & 53444 & 8.3 & 343.5 & 15904 & 1.5 & 2668\\ 
$\tau_f^h$ & s & 59.3 & 7372 & 163.1 & 321.2 & 2222 & 302.4 & 668.2 & 452.4\\
$\ell_p$ & mm & 1.445 & 4.344 & 0.002 & 7.787 & 1.309 & 0.004 & 17.97 & 0.007\\ 
$K$ & 1 & $1.38{\rm x}10^{4}$ & 13.0 & $1.29{\rm x}10^{6}$ & 231 & 3.9 & $6.18{\rm x}10^{4}$ & 3.4  & 1162  \\
$\tau_{\rm{plm}}$ & s & 1.8 & 5.3 & 1.9 &  1.3 & 3.7 & 1.3 & 0.7 & 0.7 \\
$\tau_{\rm{plt}}$ & s & $1.2{\rm x}10^{-3}$ & $4.1{\rm x}10^{-1}$ & $6.7$ & $5.2{\rm x}10^{-4}$ & $3.1$ & $8.3$ & $7.1{\rm x}10^{-4}$ & $1.4{\rm x}10^{1}$    \\
\vspace{-0.3cm}
 & & & & & & & & & \\
$\frac{\tau_{\rm{plt}}}{\tau_{\rm{plm}}}$& 1 & $6.5{\rm x}10^{-4}$ & $7.7{\rm x}10^{-2}$ & $3.5$ &  $4.1{\rm x}10^{-4}$ & $8.4$ & $6.2$ & $9.9{\rm x}10^{-4}$ & $2.1{\rm x}10^{1}$\\
\hline
\end{tabular}
\end{center}
\end{table*}

From the \textbf{theoretical/numerical} point of view, within the OB approximation and assuming that the flow is incompressible, the 3D RBC is fully described by well-known equations of motion ~\cite{AhlersRevModPhys,ChillaSchumacher} which, however, must be complemented with the boundary conditions (BC's). While for the velocity field the no-slip BC \cite{NoSlipVel} on all inner surfaces of the RBC cell are taken as justified, the BC's for the temperature field can be expressed, for example, in the Dirichlet form -- constant temperature (CT) of the solid-fluid boundary or as a constant heat flux (CF) supplied via entire area of the bottom of the RBC cell~\cite{VerziccoSreeni,VerziccoLohse}. 

In \textbf{experimental} studies of RBC, however, the temperature BC's at the interface between the fluid and the plate are always a combination of CT and CF. The main motivation of our study is that their relative weight can be, up to some degree, experimentally adjusted \textit{in situ}. 
There are many experimental parameters that affect the RBC flow under study. Following our earlier work \cite{OurNewJPhys}, we could loosely divide them into two groups: geometrical and physical.
The first group includes the actual size and shape of the cell (e.g., rectangular or cylindrical), thickness of walls and plates, their surface roughness or possible deviation from the horizontal position. The second group includes the actual physical properties of the working fluid as well as of construction materials of the RBC cell, such as thermal conductivity and heat capacity of plates($\lambda_p; c_p$) and walls, the thermal conductivity of the electrical leads and, generally, the physical properties of the surrounding medium. Although various approaches to correct the raw data with respect to finite thermal conductivity of plates~\cite{ChillaCastaing} and walls~\cite{RocheWalls}, parasitic heat leaks, adiabatic thermal gradient, thermal radiation~\cite{ThRadiation} or non OB effects~\cite{nonOBWuLib,OurNewJPhys,nonOBSkrUrban, Pandey} have been attempted by various authors, it is generally very difficult if not impossible to fully eliminate all these factors. In order to single out and appreciate the role of BC on the RBC flow under study, it therefore seems the best to perform the experiment under the same conditions while changing the BC only.

On the other hand, we believe that it is instructive to compare, for selected RBC flows fully described by Ra, Pr and $\Gamma=1$, also additional parameters relevant for a typical RBC experiment assuming it performed in the same cell. We have chosen three typical data points measured in this study with cryogenic He gas (see  Table~\ref{tab.1}) that belong to ranges of Ra obeying power scaling ${\rm{Nu=Nu(Ra)} \propto {\rm{Ra}}^\gamma} $ ($\gamma \approx 2/7$, crossover regime and $\gamma \approx 1/3$ \cite{OurNewJPhys,OurPRL11})  
and compare them with complementary hypothetical 
turbulent RBC flows, assuming them generated in the same cell at ambient temperatures using frequently used working fluids: $\rm{H_2O}$ and ${\rm{SF}}_6$. 
The numerical value $\alpha_f \Delta T  < \approx 0.2$ is conventionally understood as a satisfactory OB criterion. The thickness of the thermal boundary layer $\ell_{\rm{BL}}=L/(2\rm{Nu})$ is naturally the same for the same Nu, 
however, the heat currents $Q_b = q S_p$ ($S_p$ being the plate area) required to be applied to the bottom plate are for complementary RBC flows very different. It would take the time $\tau_p^h$ to heat just the bottom plate alone (assuming it thermally isolated) by $\Delta T$, and time $\tau_f^h$ to heat the working fluid by $\Delta T/2$, i.e., to the temperature of turbulent bulk of the RBC flow. An important factor is the heat conductivity, $\lambda_p$, of the plates. Its influence on Nu was thoroughly studied \cite{ChillaCastaing,Verzicco2004} and experimentally confirmed by Brown \textit{et al.}~\cite{Brown}, who used $\rm{H_2O}$ in otherwise identical RBC cells with Cu ($\lambda_p \approx 391$~Wm$^{-1}$K$^{-1}$) and Al ($\lambda_p \approx 161$~Wm$^{-1}$K$^{-1}$) plates and concluded that low $\lambda_p$ appreciably diminishes the heat transport efficiency, at least in RBC cells of size similar to our own~\cite{BrnoCellRSI}.  An important requirement is that the ratio $\lambda_p/({\rm{Nu}} \lambda_f)$ is high, which for Cu plates and water is low (see Table~\ref{tab.1}) but even lower for Al plates. Note that for ${\rm{SF}_6}$ and especially for cryogenic He this ratio is about two (three) orders of magnitude higher.

The key role for establishing the ratio of CT versus CF BC's is played by thermal plumes. Let us consider a typical plume: a two-dimensional sheet-like structure of temperature $\approx T_b$ (hot plume) or $\approx T_t$ (cold plume) and thickness comparable to 
$\ell_{\rm{BL}}$, which initially extends in the vertical direction, eventually to be bent by LSC. If such a plume of area $S$ abruptly detaches, it takes with it (leaves behind) heat $Q_p \approx S \ell_{\rm{BL}} c_f \rho_f \Delta T/2$, equivalent to a thermal hole in the plate, of thickness $\ell_p \approx 2Q_p/(
S\Delta T c_p \rho_p)$ (see  Table~\ref{tab.1}); this thermal hole must be refilled using the heat flux delivered by a heater via thermal conduction of the plate. From this point of view, an important parameter is $K = (\rho_p c_p \lambda_p)/(\rho_f c_f Nu \lambda_p)$ \cite{Schlichting}. 
 The characteristic time between two successive plumes has been estimated by Castaing \textit{et al.}~\cite{Castaing89} as $\tau_{\rm{plm}}= {\rm{(Ra Pr)^{1/2}}}/(4 {\rm{Nu}}^2)$. It decreases with Ra, since Nu increases faster than Ra$^{1/4}$. To assure CT BC the plate should be fast enough to provide consecutive plumes with enough heat by thermal diffusion, which occurs within a characteristic time  $\tau_{\rm{plt}}=({\rm{RaPr}})^{1/2}(e/L)^2(\kappa_f/\kappa_p)$~\cite{Verzicco2004,Castaing89}. This means that CT BC will be better achieved if $K$ is big and the ratio $\tau_{\rm{plt}}/\tau_{\rm{plm}}$ is small, which is out of the three considered cases best achieved for cryogenic He.

\begin{figure}[t]
\centering
\includegraphics[width=0.945\linewidth]{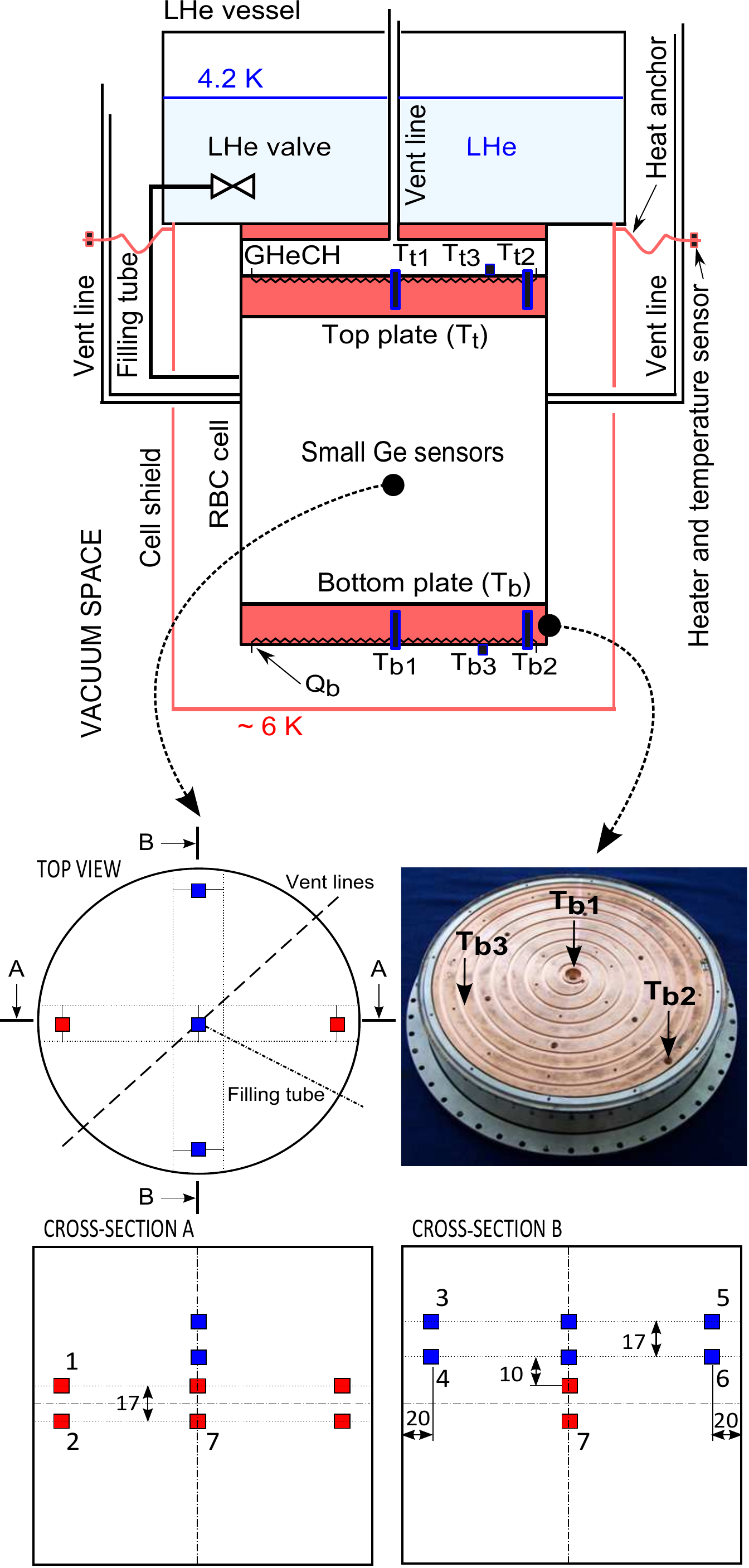}
\caption{The sketch of the Brno RBC cell. From the top plate, most of heat is removed via the He gas heat exchange chamber (GHeCH) to liquid He vessel above it. $T_t$ is roughly set by pressure in the GHeCH and more precisely by the distributed heater. Note the positions (distances in mm) of small Ge sensors (numbered are those used in this study) in the cell interior, the finely calibrated Ge sensors $\rm {T}_{t1}, \rm {T}_{t2}, \rm {T}_{b1}$ and $\rm {T}_{b2}$ embedded in the plates and two fast diodes $\rm {T}_{t3}$ and $\rm {T}_{b3}$ at the outer surfaces of plates (see the detail of the bottom plate, showing the spiral grove where the heater is glued, delivering approximately uniformly both the heat flux $q$ and the PID-control heat flux aimed to stabilize $T_b$. }
\label{fig:BrnoCell}
\end{figure}

\textbf{Experiment.}  In order to appreciate the role of BC on RBC flow, we perform the experiment under nominally the same conditions while switching on and off the PID-stabilizing scheme of the bottom plate temperature $T_b$. We use the updated version of the Brno experimental cell~\cite{BrnoCellRSI}, shown in Fig.~\ref{fig:BrnoCell}. Essential improvements are the following: (i) the original mid flanges on the sidewall have been gradually deformed in previous experiments and found prone to leakage at high pressure of the working fluid; these flanges were therefore replaced and the joints welded together; (ii) several small Ge temperature sensors (Ge-on-GaAs film resistance thermometers, \cite{Mitin}) attached to tightly stretched thin constantan wires have been installed, via newly made sidewall feedthroughs;
their geometrical positions are shown in Fig.~\ref{fig:BrnoCell}; and (iii) in addition to the precisely calibrated stable Ge sensors embedded in the plates, fast DT-670 Silicon Diodes (Lake Shore) have been attached to both plates. 

\begin{figure}[t]
\begin{center}
\includegraphics[width=1.0\linewidth]{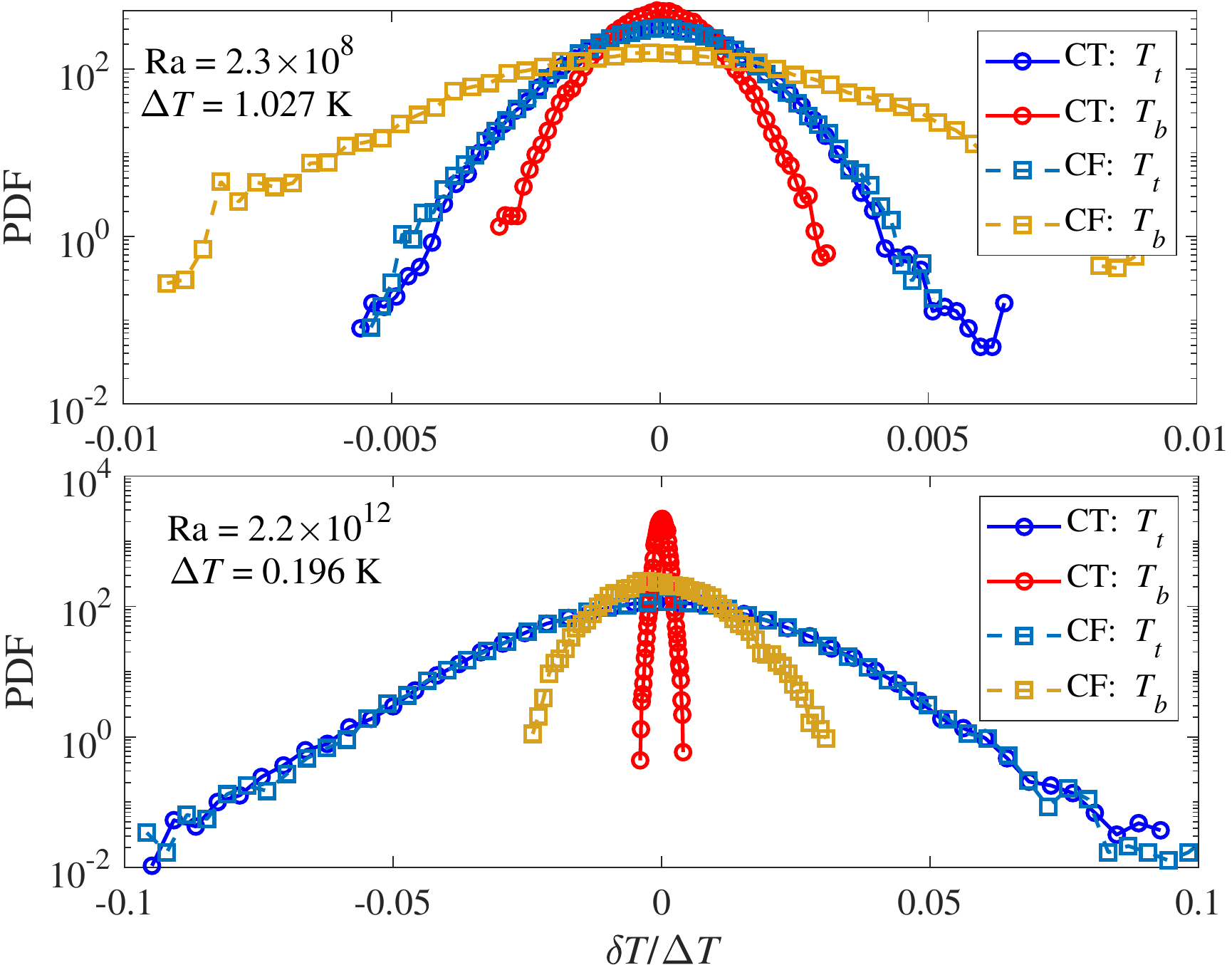}
\end{center}
\caption{Examples of the PDFs of the temperatures $T_b(t)$ and 
$T_t(t)$  fluctuating about mean temperatures $\langle T_b\rangle$ and $\langle T_t\rangle$, measured by fast responding diodes $\rm{T_{b3}}$ and $\rm{T_{t3}}$ placed at the outside horizontal surfaces of the bottom and top plates at Ra as indicated, for two sets of BC discussed in the text. The PID control of the bottom plate temperature results in significantly narrower PDFs (red circles), in comparison with ``standard" CF heating (orange squares), while the PDFs measured at the top plate (blue symbols) remain unaffected.}
\label{fig:pdf}
\end{figure}

Due to rather complex thermal connection of the top plate with the liquid helium vessel above it, partly via a stainless steel sidewall but mainly via the He gas heat exchange chamber (GHeCH) which itself represents a convection cell, we have focused on changing BC at the bottom plate and compare two distinctly different cases. In both of them, the heat is supplied to the bottom plate via a distributed wire heater. 
As the distance between heater turns is smaller than the plate thickness, the heat delivered to its upper surface, in the absence of convective flow in the RBC cell, can be thought of as steady and uniformly distributed. Turbulent RBC flow breaks this symmetry both in time and space. Although the total heat flux delivered by the resistive heater to the outer side of the bottom plate remains constant, due to thermal plumes detachment and dynamical thermal properties of the bottom plate, the CF BC is not strictly valid  
at the bottom solid-fluid boundary of the RBC flow. Despite this caveat, also in view of numerical studies such as \cite{JohnstonDoering}, hereafter we call this Case 1 as CF BC. 

We note that delivering constant heat flux (CF) to the outer side of bottom plate while controlling the mean top plate temperature (via adjusting the pressure in the exchange chamber and, additionally, by fine tuning via uniformly distributed resistive heater glued in the spiral grove on the upper surface of the top plate achieved by using a PID control) is the ``standard" way of generating statistically steady turbulent RBC flows studied in our previous experiments \cite{OurNewJPhys,OurPRL11,OurJFM2017,OurPRE19} and references therein. 

Case 2 to compare with, hereafter called CT BC, differs in that the bottom plate heater is included in the PID control feedback loop, designed to keep the temperature of the bottom plate stable. The PID scheme uses the reference signal from the fast-responding diode $\rm {T}_{b3}$.
In both cases, the mean temperature difference $\Delta T = \langle T_b\rangle-\langle T_t\rangle$ is kept constant, where the mean temperatures $\langle T_b\rangle$ and $\langle T_t\rangle$ are accurately determined by finely calibrated Ge sensors $\rm {T}_{b1}$, $\rm {T}_{b2}$ and $\rm {T}_{t1}$, $\rm {T}_{t2}$.
The fluctuating values  $T_b(t)$ and $T_t(t)$ are monitored by home-calibrated diodes $\rm {T}_{b3}$ and $\rm {T}_{t3}$, and the temperature fluctuations $T_N(t) (N=1...12)$ in various places of the cell interior by small Ge-on-GaAs film sensors \cite{Mitin}.

\begin{figure}[t]
\begin{center}
\includegraphics[width=1.0\linewidth]{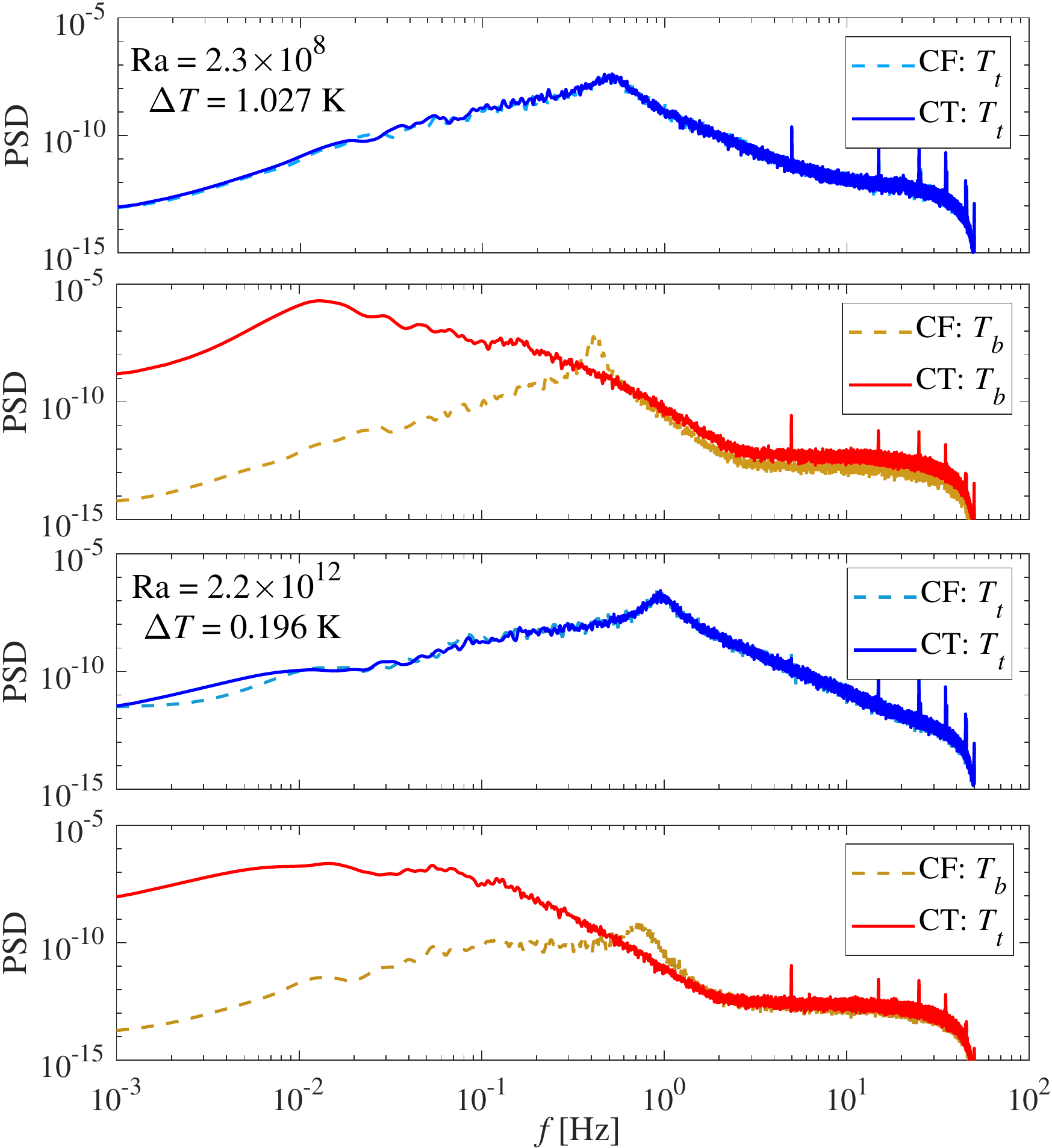}
\end{center}
\caption{PSDs of the fluctuating temperature calculated for the cases shown in Fig.~\ref{fig:pdf} display significant depletion at frequencies below 0.4 Hz (0.7 Hz) for the lower (higher) Ra cases; faster temperature fluctuations of the bottom plate are hardly affected by the PID control (red lines). At the top plate, PSDs (blue lines) remain entirely unaffected at all measured Ra. }
\label{fig:PSD}
\end{figure}

\begin{figure}[t]
\begin{center}
\includegraphics[width=\linewidth]{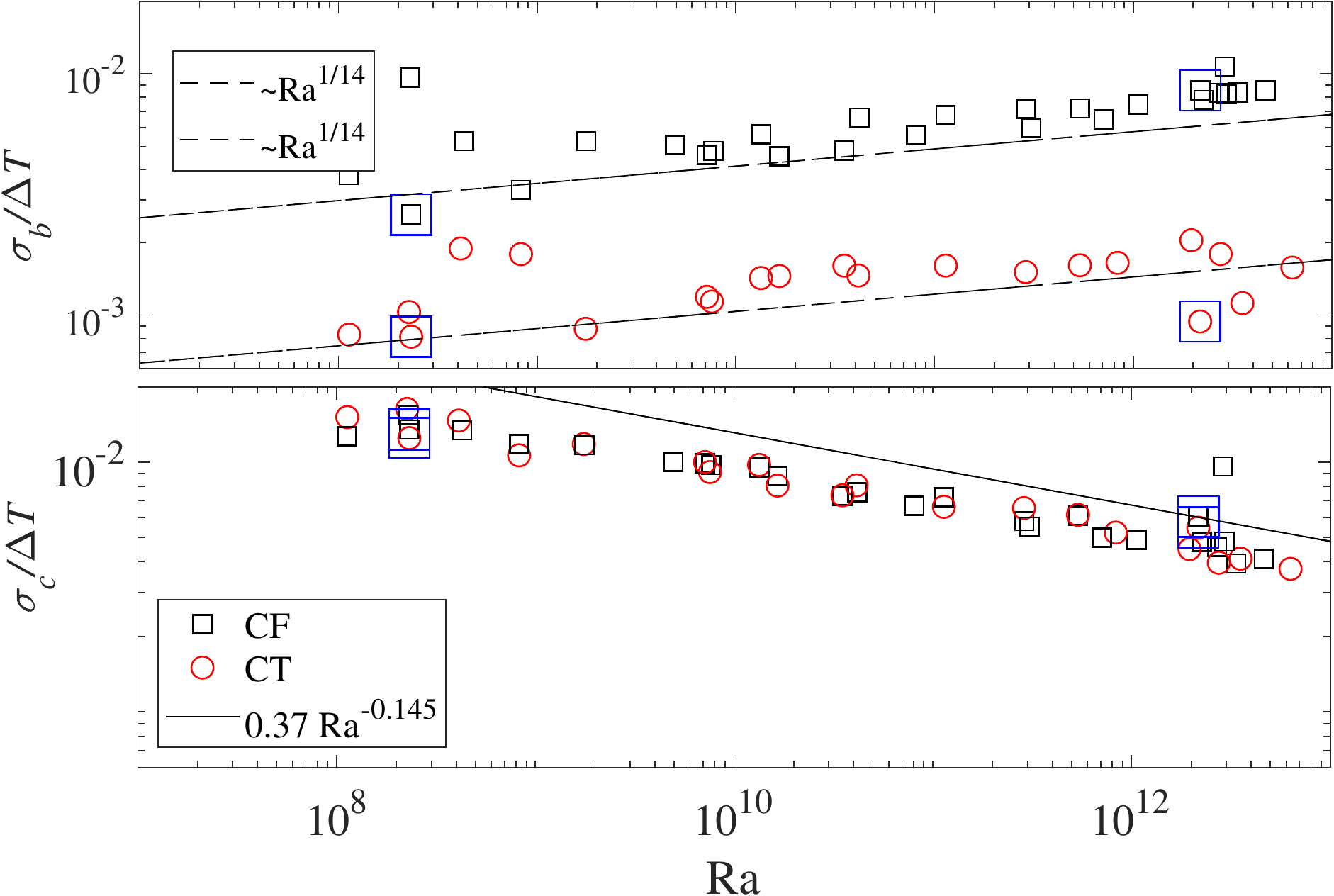}
\end{center}
\caption{Top panel shows the mean value of the temperature fluctuations $\sigma_b$ measured by a diode $\rm {T}_{b3}$ mounted on the outer side of the bottom plate, normalized by $\Delta T$, plotted versus Ra for CF BC (black open squares) and for the case of PID-controlled $T_b$; CT BC (red open circles). The dashed lines indicate the slope of ${\rm{Ra}}^{1/14}$. Bottom panel displays the mean of the temperature fluctuations  $\sigma_c/\Delta T$ measured by a small Ge sensor No 7 in the centre of the RBC cell for CF BC (black open squares) and for the case of PID-controlled $T_b$; CT BC (red open circles). The solid line represents the the best fit through the same data measured by Niemela \textit{et al.} \cite{NatureNiemela}: $\sigma/\Delta T= 0.37 {\rm Ra}^{-0.145}$. Large blue square symbols highlight cases shown in Figs. 2 and 3.}
\label{fig:fluct}
\end{figure}

The probability density functions (PDFs) of the temperature fluctuations of the plates $T_b(t)$ and $T_t(t)$ are evaluated  using the signal from the fast-responding diodes $\rm{T_{b3}}$ and $\rm{T_{t3}}$ (see Fig.~\ref{fig:BrnoCell}) with and without the PID control of the bottom plate temperature. All measured PDFs of the fluctuating $T_b(t)$ and 
$T_t(t)$ about mean temperatures $\langle T_b\rangle$ and $\langle T_t\rangle$ are approximately symmetric and of Gaussian shape, see Fig.~\ref{fig:pdf}.  We have chosen two examples of Ra belonging to different ${\rm{Nu(Ra)}} \propto {\rm{Ra}}^\gamma$ scaling, the lower one in the range of $\gamma \approx 2/7$; the upper one above the crossover to $\gamma \approx 1/3$ \cite{OurPRL11,OurNewJPhys}. For all investigated Ra, the PID control of the bottom plate temperature results in significant narrowing of the bottom plate PDFs, while the PDFs measured at the top plate are not appreciably affected. 

It is instructive to calculate and compare the power spectral density (PSD) of temperature fluctuations for the PID control on and off. As shown in Fig.~\ref{fig:PSD}, the PID control results in significant depletion of PSDs at low frequencies below about $0.4-0.7$ Hz, while faster temperature fluctuations of the bottom plate are hardly affected. The top plate PSDs remain at all measured Ra entirely unaffected.

The top panel of Fig.~\ref{fig:fluct} displays the mean value of the temperature fluctuations $\sigma_b$ measured by a diode $\rm{T_{b3}}$ normalized by $\Delta T$, plotted versus Ra. While $\sigma_b /\Delta T$ slightly increases with increasing Ra (approximately $\propto {\rm{Ra}}^{1/14}$) for both CF and CT BC on the bottom plate, the imposed CT BC reduces its numerical values by a factor of about four. The same quantity, $\sigma/\Delta T$ in the top and bottom Cu plates in rectangular RBC cells of various sizes was measured at ambient temperatures under CT and CF BC in a similar study by Huang \textit{et al.}~\cite{HuangXia}, by using $\rm{H_2O}$ as the working fluid. It is remarkable that Fig.~1d of Ref.~\cite{HuangXia} clearly shows, for both the top and bottom plates, the opposite tendency in the $\sigma/\Delta T$ versus Ra dependence. We speculate that this apparent discrepancy could be explained by very different dynamic characteristics of cryogenic He and ambient temperature $\rm{H_2O}$ turbulent RBC experiments, as some of them differ by orders of magnitude - see Table~\ref{tab.1}.
  
\begin{figure}[t]
\centering
\includegraphics[width=1.0\linewidth]{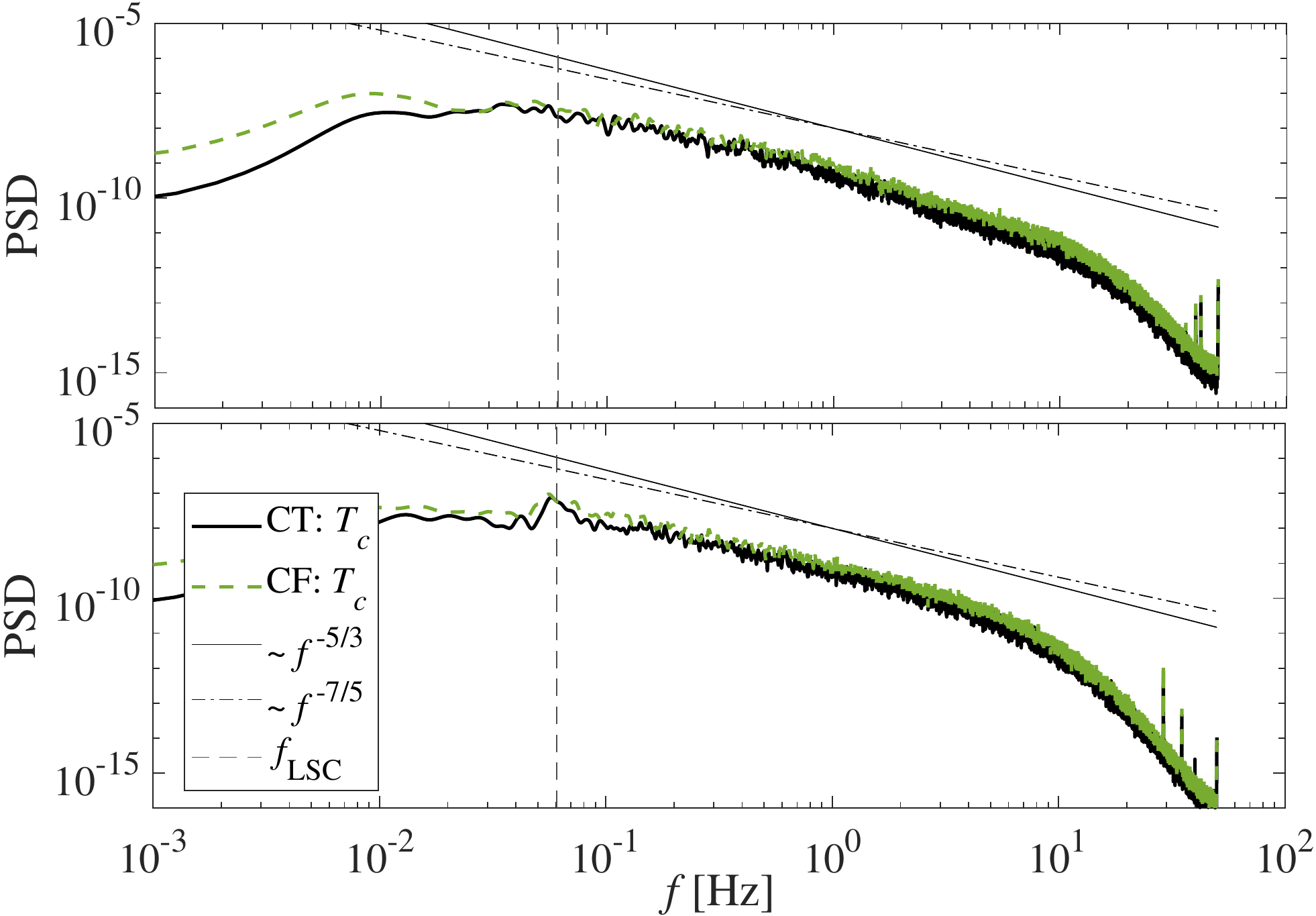}
\caption{PSDs of the temperature fluctuations in the bulk RBC at Ra= $2.2 \times 10^{12}$ for CF (dashed green lines) and CT-like (solid black lines) BC at the bottom plate. The PSD in the top panel are measured by sensor No 7 in the geometrical centre of the cell, those in the bottom panel, displaying the LSC peak at 0.06 Hz, by sensor No 1, 20 mm from the sidewall. The lines represent the slopes of Bolgiano and Obukhov-Corrsin scaling with, respectively, -7/5 and -5/3 roll-off exponents.  }
\label{fig:BulkSpectra}
\end{figure}

Let us now discuss the main issue of this study: what changes, if any, are experimentally observed in the \textit{bulk }of the RBC flow as a consequence of distinctly different BC at the bottom plate. We start with the same 
quantity, $\sigma/\Delta T$, but measured now not at the plates but in the centre of the RBC cell.  The bottom panel of Fig.~\ref{fig:fluct} shows that, contrary to the situation at the bottom plate, $\sigma_c/\Delta T$ in the centre is not appreciably sensitive to the change of BC at the bottom plate and scales $\propto {\rm{Ra}}^{-1/7}$, and behaves in accord with our previous studies~\cite{OurJFM2017} performed in the RBC Brno cell as well as with the seminal work of Niemela \textit{et al.}~\cite{NatureNiemela} quoting the best fit $\sigma/\Delta T= 0.37 \mathrm{Ra}^{-0.145}$, shown in the bottom panel of Fig.~\ref{fig:fluct} as a solid line for comparison. 

Fig.~\ref{fig:BulkSpectra} shows examples of PSDs of the temperature fluctuations in the centre of the RBC cell (top) and at the mid plane 20 mm from the sidewall (bottom) measured at ${\rm{Ra}}=2.2 \times 10^{12}$. As it is typical for confined high Ra RBC flow, the PSD measured near the sidewall displays the LSC peak, in this case  at 0.06 Hz, which is used to calculate the mean velocity of the LSC, the ``wind". In accord with \cite{NatureNiemela}, the PSD are consistent with a roll-off rate of $-7/5$ for low frequencies where Bolgiano scaling seems appropriate, whereas for higher frequencies, the classical Obukhov-Corrsin scaling with the roll-off exponent $-5/3$ appears more appropriate. 
The key observation is that at all investigated Ra, except for slight depletion at very low frequencies below 0.02 Hz, the bulk PSD of the temperature fluctuations are unaffected by the imposed change of BC on the bottom plate. 

\begin{figure}[t]
\begin{center}
\includegraphics[width=1.0\linewidth]{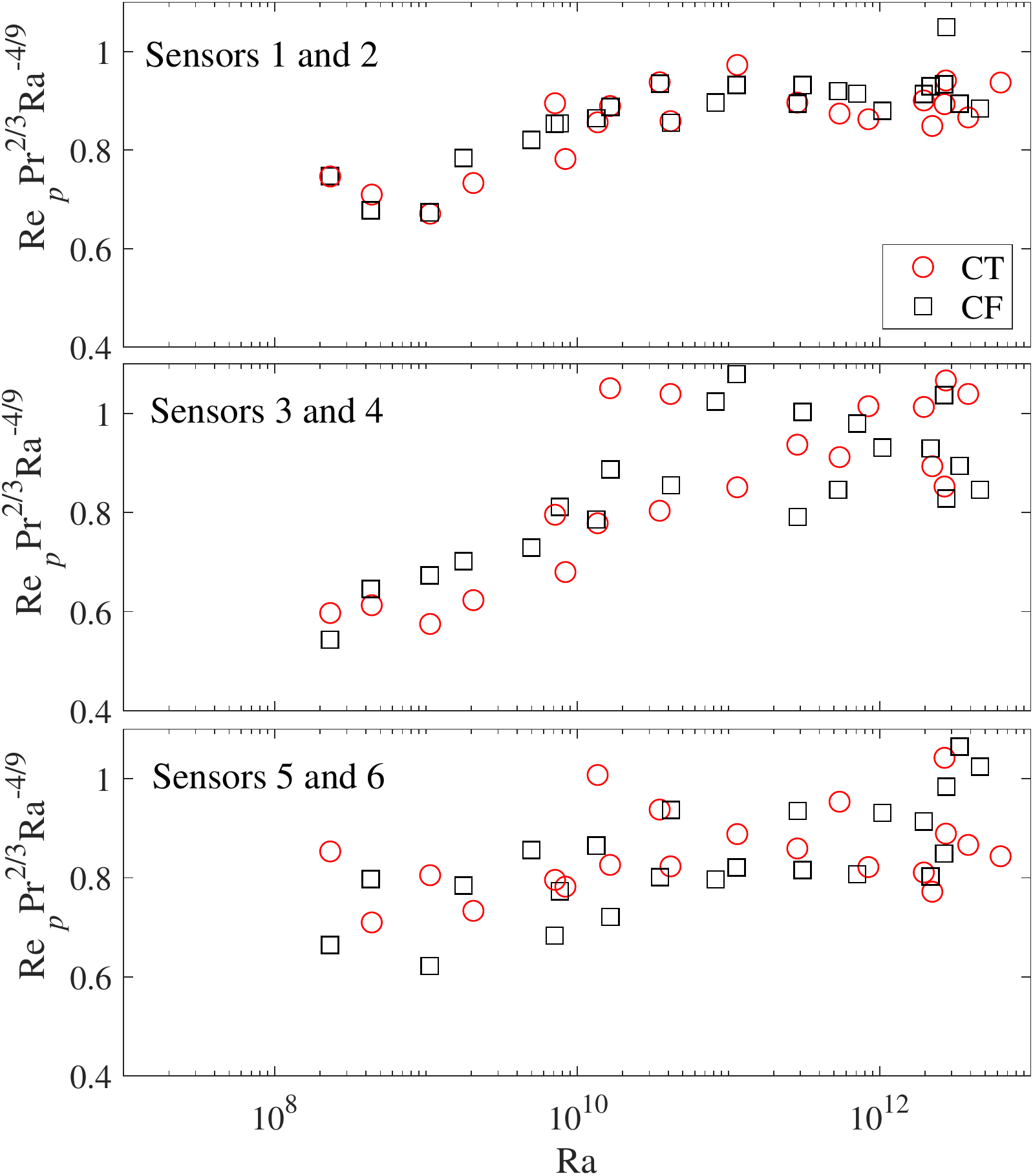}
\end{center}
\caption{Reynolds number $\rm{Re}_{\it p}$ calculated using temperature signals from pairs of vertically spaced Ge sensor as indicated, plotted in a compensated form versus Ra.}
\label{fig:Reynolds}
\end{figure}

\begin{figure}[h!]
\begin{center}
\includegraphics[width=\linewidth]{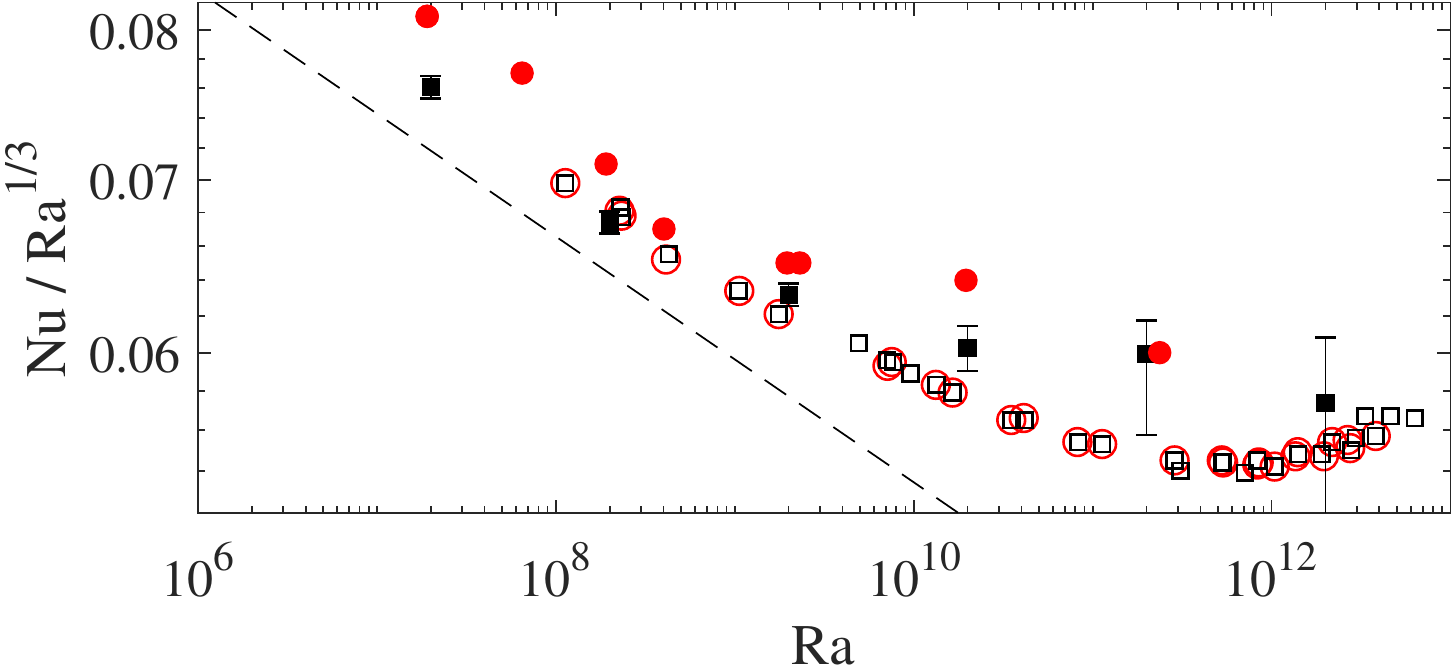}
\end{center}
\caption{Compensated Nu plotted versus Ra for ``standard" RBC generation (CF BC on the bottom plate, open squares) and for the case of PID  control of temperature of the bottom plate (BC closer to CT, open red circles). The data series have been measured under the same experimental conditions and only the same basic corrections, due to adiabatic gradient and parasitic heat leak have been applied. Also shown are CF (filled black squares) and CT (filled red circles) numerical data for $\Gamma=1/2$ and Pr=0.7 \cite{VerziccoLohse}, see the text for details. The dashed line indicates the slope ${\rm{Nu}}\propto {\rm{Ra}}^{2/7}$.}
\label{fig:NuRa}
\end{figure}

In our previous work \cite{OurJFM2017} we discussed in detail several definitions of Reynolds numbers, evaluated them using the ``standard" CF BC and directly compared them with results published by other authors. Here we utilize temperature fluctuations measured by various single and pairs of Ge sensors in the cell (see Fig.~\ref{fig:BrnoCell}) and compare Reynolds numbers and their scaling with Ra for CF and CT BC at the bottom plate. 

In the case of one probe measurement, the characteristic frequency $f_0$ determined from the peak of the near-wall PSD of the temperature fluctuations (an example shown at the bottom of Fig.~\ref{fig:BulkSpectra}) is used in definition of the frequency-based Reynolds number ${\rm{Re}}_{f0}= 2 L^2 f_0/\nu_f$. We already discussed that the near-wall PSD remain unaffected by the change of the BC at the bottom plate, so ${\rm{Re}}_{f0}$ is unaffected, too. 

In the case of two probe measurements, the simplest approach relies on the Taylor's frozen flow hypothesis and uses the time delay between temperature fluctuation records at two nearby sensors spaced by a vertical distance $d$, which determines the mean velocity $U_p$. The corresponding Reynolds number is defined as ${\rm{Re}}_{p}= L U_p/\nu_f$. In  \cite{OurJFM2017} we claimed observation of a crossover in the slope of  ${\rm{Re}}_{p}$ with Ra around $10^{10}$ (complementary to the crossover in ${\rm{Nu(Ra) \propto Ra}}^\gamma$ scaling from $\gamma \approx 2/7$ to $\gamma \approx 1/3$). We confirm this crossover for both CF and CT BC on the bottom plate; it is clearly seen in the compensated plot of ${\rm{Re}}_{p} {\rm{Pr}}^{2/3}/ {\rm{Ra}}^{4/9}$ versus Ra, displayed in the top panel of Fig.~\ref{fig:Reynolds}. \michal{We have evaluated Reynolds numbers according to all definitions discussed in our previous work \cite{OurJFM2017} and found them hardly affected by the changing CF and CT BC at the bottom plate. The ${\rm{Re}}_{p}$ data shown in the top panel of Fig.~\ref{fig:Reynolds} were evaluated using sensors 1 and 2 (see Fig.~1) which in all experimental runs displayed very rare reversals \michall{of the LSC direction: $1.5 \pm 0.2$ ($1.8 \pm 0.3$) reversals$/$hour for CF (CT) cases, as the sensors presumably lied near the main LSC plane. The data in the middle and bottom panels are from the sensor pairs 3, 4 and 5, 6, which lied in the plane perpendicular to the previous pair and experienced more frequent reversals of an auxiliary flow: $9.7 \pm 0.4$ ($9.9\pm 0.4$) reversals$/$hour for CF (CT) cases. Similar situation was observed by Sun {\it et al.}~\cite{SunXiaTong} using PIV combined with thermometry. More detailed statistical study of LSC reversals as well as analysis employing the so-called elliptic approximation in evaluation of $\rm{Re}$ will be published elsewhere.}}

Last but not least we now discuss the essential feature of turbulent RBC flow - its ability to transfer heat, usually expressed in
dimensionless form, by the Nusselt number. The key question is: Does Nu depend on boundary conditions? Our experimental answer is provided in a graphical form in Fig.~\ref{fig:NuRa}: Changing CF to CT BC at the bottom plate does not appreciably change the heat transfer efficiency, at least over the investigated range $10^8 < {\rm{Ra}} < 3 \times 10^{12}$. Being fully aware of the fact that accurate determination of the Nu(Ra) dependence involves application of various corrections to the raw data, we do not claim here the absolute accuracy of the displayed compensated Nu(Ra) dependence. We stress, however, that the only difference between the displayed two sets of data is the \textit{in situ} change of CF (or rather CF-like) and CT (or rather CT-like) BC at the bottom plate as discussed in detail above.

This experimental result can be compared with complementary numerical studies. Following the earlier simulations of  Amati \textit{et al.}~\cite{Amati}, Verzicco \& Sreenivasan \cite{VerziccoSreeni}, and 2D simulations of Johnston \& Doering \cite{JohnstonDoering}, Stevens, Lohse \& Verzicco \cite{VerziccoLohse}  performed thorough 3D simulations with improved accuracy, under unconditional validity of the OB approximation for $\Gamma=1/2$ and Pr=0.7. Fig.~\ref{fig:NuRa} shows their direct numerical simulation data with CT and CF BC at the bottom plate (with CT BC at the top plate in both cases); only the data obtained on identical grids are displayed in order to rule out resolution effects in calculations as much as possible. In the simulations with CF BC at the bottom plate they calculate Nu from the average $q$ at plates. The error bars are larger in CF than in CT BC because it took more time to reach the statistically steady state. Up to about $10^{10}$ in Ra, CF BC generally results in slightly lower values of Nu (which is not seen experimentally). The difference between CF and CT simulations is, however, rather small, seems to decrease with increasing Ra and disappears within the (statistical) error above $10^{11}$ in Ra. 

While simulations \cite{VerziccoLohse} are undoubtedly relevant to our experiment, they do not take directly into account possible influence of physical and geometrical properties of the RBC cell, especially of its plates. This was attempted in the recent study of Foroozani, Krasnov and  Schumacher~\cite{ForoozaniSchumacher}, who studied numerically the confined RBC flow bounded by two copper plates from above and below and applied CT and CF BC at the upper surface of the top plate and at the lower surface of the
bottom plate (see Fig. 1 in Ref.~\cite{ForoozaniSchumacher}), requiring the continuity of temperature and heat flux at the solid-fluid interfaces. This configuration is denoted as the conjugated heat transfer case.
The authors found that for fixed Pr and Ra in the range $10^7-10^8$ the turbulent heat
transfer is \textit{enhanced} for the CF and conjugate heat transfer cases in comparison to the standard CT case. It will be interesting to extend these investigations which take into account various geometrical and physical properties of plates and working fluids, such as those suggested in Table~\ref{tab.1}. This approach should shed new light on experimentally found differences observed when performing experiments using different working fluids in RBC cells of the same shape with plates made of different materials of strongly temperature dependent physical properties. 

\textbf{To conclude}, we have performed an experimental study aiming to clarify the role of BC on high Ra turbulent RBC flow. Our findings can be summarized as follows. Changing the CF-like BC to CT-like BC on the bottom plate by employing the PID control of $T_b$ results in (i) significant narrowing of temperature PDF and suppression of the low frequency part of the PSD evaluated with the use of a fast-responding sensor attached to the bottom plate itself, but hardly affects (ii) the dynamic thermal characteristics of the top plate, (iii) PSD of temperature fluctuations measured in the centre of the cell as well as in the middle plane near the sidewall, \michall{ (iv) the rate of direction reversals of large scale circulation and finally (v) the Nu(Ra) and (vi) Re(Ra) scaling}. 

\acknowledgments
We thank R. Verzicco and  R.J.A.M. Stevens for providing the simulation Nu(Ra) data  \cite{VerziccoLohse} in a digital form and to V. Musilov\'{a} and K.R. Sreenivasan for stimulating discussions. The authors acknowledge support of Czech Science Foundation under GA\v{C}R 20-00918S. \michal{M.M. acknowledges funding by EU structural and investment funds and MEYS CR OPVVV CZ.02.2.69/0.0/0.0/18 070/0009944.}

\end{document}